\documentclass[a4paper,12pt]{article}
\usepackage{epsfig}
\newcommand{\myskip}{\vspace{\baselineskip}}
\newcommand{\mysection}[1]{\par\myskip\noindent\textbf{#1}\myskip\par}

\begin{document}
\begin{center}
{\large\textbf{Few-Electron Quantum Dots and Disks in Zero Magnetic Field: Possible Indications on a Liquid-Solid Transition}} 

\myskip 
S. A. Mikhailov\\
Institute for Physics, University of Augsburg, D-86135 Augsburg, Germany

\myskip 




\begin{abstract}
Exact-diagonalization studies of few-electron quantum dots and disks are performed, with the aim to investigate a Wigner cluster -- Fermi liquid crossover in zero magnetic field at varying strength of Coulomb interaction. A clear indication of a transition of a liquid-solid type in the ground state is found in a more adequate quantum-disk model.
\end{abstract}

\end{center}

\mysection{Introduction}
A number of exact-diagonalization [1-3], 
quantum Monte Carlo [4-7], 
and other [8] 
studies of few-electron quantum dots in zero magnetic field were recently performed, with the aim to investigate the crossover Fermi liquid -- Wigner cluster in these small interacting electron systems. Detailed knowledge of the physics of such a crossover could be compared with that obtained for macroscopic two-dimensional electron systems (2DES) [9-10] 
and might shed light on the nature of the metal-insulator transition in 2DES [11]. 
However, apart from the case of a quantum-dot helium [1], 
a full understanding of the energy spectra and physical properties of an $N$-electron quantum dot was not yet achieved, and results obtained by different methods often contradict each other. 

In this work I present a number of results which may contribute to these continuing efforts. I perform an exact-diagonalization study of a system of $N$ confined electrons as a function of the interaction parameter. Two models of the confinement are used: the traditional parabolic quantum dot model, and a more sophisticated quantum disk model. The confinement potential in the disk model is strongly non-parabolic, which leads to {\em qualitatively} different results as compared to dots. 

\mysection{Models and method}

{\sl Quantum dots.} The Hamiltonian of the system consists of the sum of the single-particle energies of $N$ electrons in the harmonic potential $V(r)=m^\star\omega_0^2r^2/2$, and the Coulomb interaction energy. I perform configuration-interaction calculations with a sufficient number of basis many-body states to achieve the convergency of the energy, and use the fact that in the limit of the infinitely strong interaction the ground state energy can be calculated exactly. The interaction parameter in the dot is given by the ratio $\lambda=l_0/a_B$ of the oscillator length $l_0=(\hbar/m^\star\omega_0)^{1/2}$ to the effective Bohr radius $a_B$. All the calculated energy levels are classified by the total angular momentum $L_{tot}$ and the total spin $S_{tot}$ quantum numbers. 

{\sl Quantum disks.} The quantum-dot model does not allow one to directly compare results with those for macroscopic 2DES since the macroscopic interaction parameter $r_s=a_0/a_B$, $a_0=(\pi n_s)^{-1/2}$, cannot be rigorously defined in dots (the density of electrons in the dot is essentially inhomogeneous; here $n_s=const$ in the density of electrons in a macroscopic 2DES). I consider another, more adequate, {\sl quantum disk} model. Given in this model is not the confining potential, but the density of the positive background $n_b(r)= n_s\tilde\theta(R-r)$, where $R$ is determined by $\pi R^2n_s=N$, and the step function $\tilde\theta(r)$ is smoothed near the edge at a distance of order of $n_s^{-1/2}$. The confining potential of the disk 
is parabolic near the origin 
and tends to zero as $
-Ne^2/r$ at $r\gg R$. The disk model ideally describes a realistic sample with a constant positive background density (and hence with a constant average electron density). In this model the parameter $r_s$ is well defined, and one can study properties of the system as a function of $r_s$ even at small electron numbers. 

The Hamiltonian in the disk model consists of the kinetic energy of $N$ electrons, and the total Coulomb energy of the system (electron-electron, electron-background and background-background interaction energies). The energy spectra and other properties are calculated in the same way as in the dot problem. 

\mysection{Results}

\begin{figure}[t]
\centering
\epsfig{file=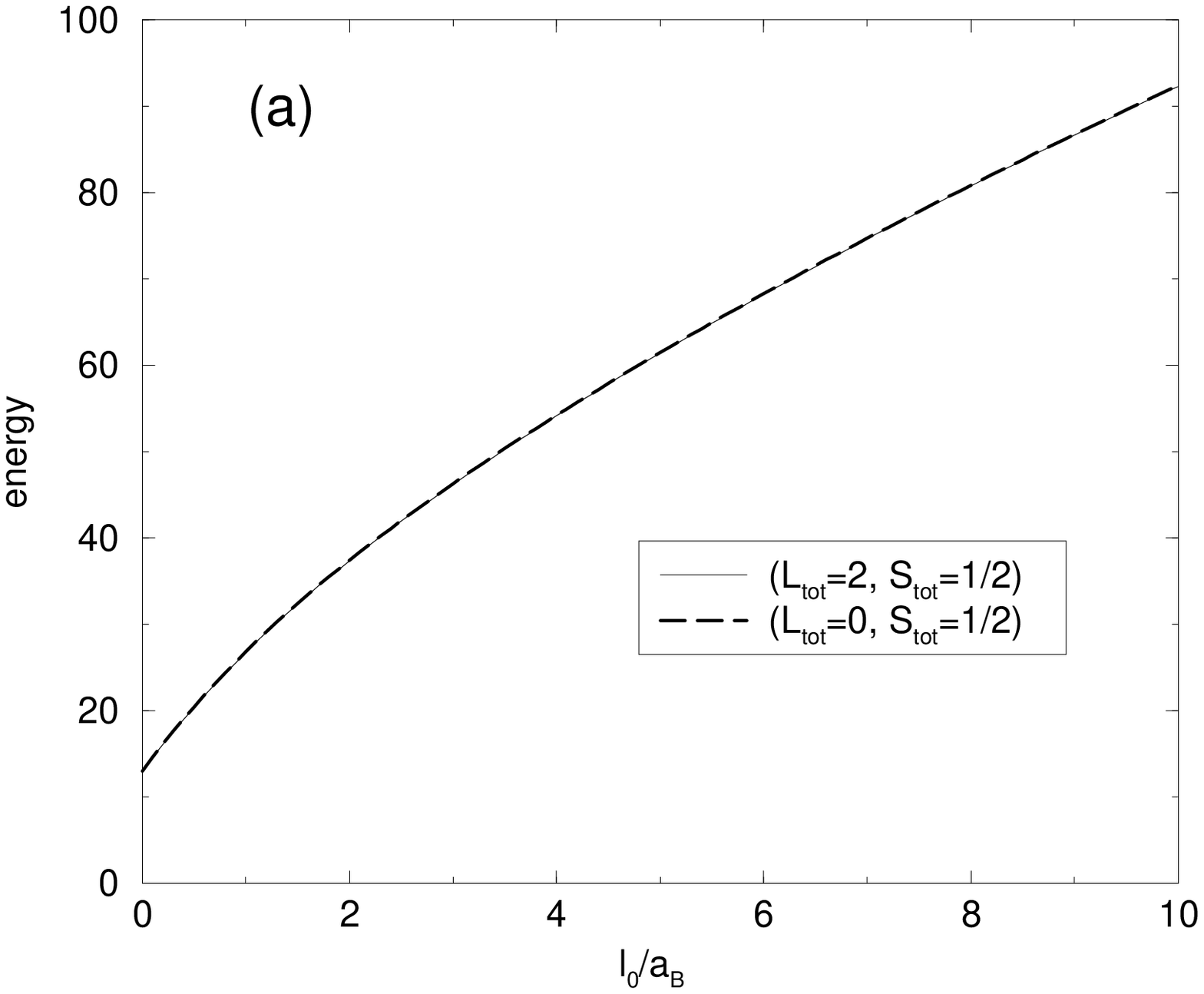,width=7.5cm}\hskip5mm\epsfig{file=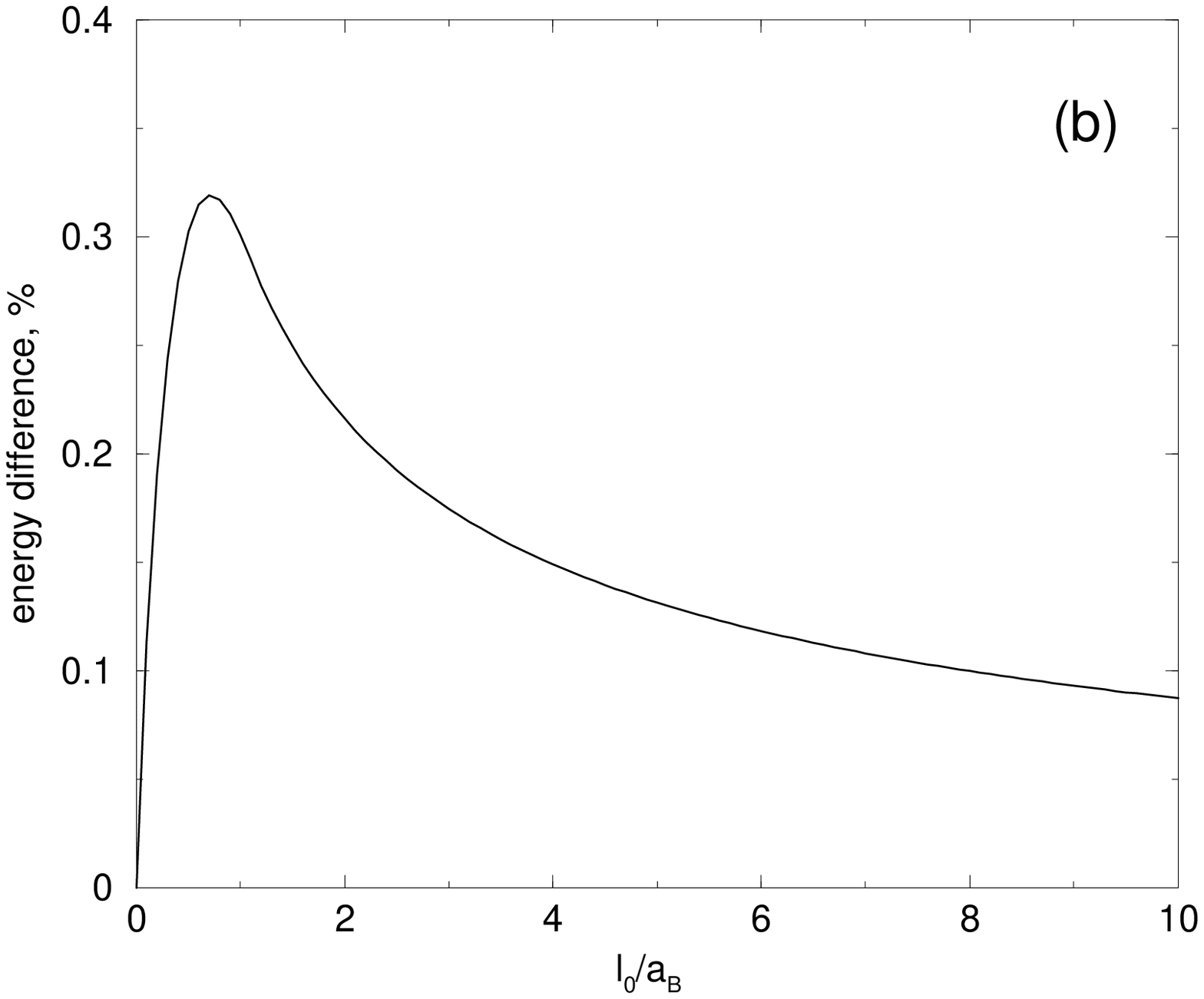,width=7.5cm}
\caption{(a) The ground $(L_{tot},S_{tot})=(2,1/2)$ and the first excited $(0,1/2)$ state energies of a 7-electron quantum dot as a function of the interaction parameter $\lambda=l_0/a_B$. The energy unit is $\hbar\omega_0$. (b) Relative difference (in \%) of these energies vs $\lambda$.}
\label{7el-energy}
\end{figure}

\begin{figure}[!h]
\centering
\epsfig{file=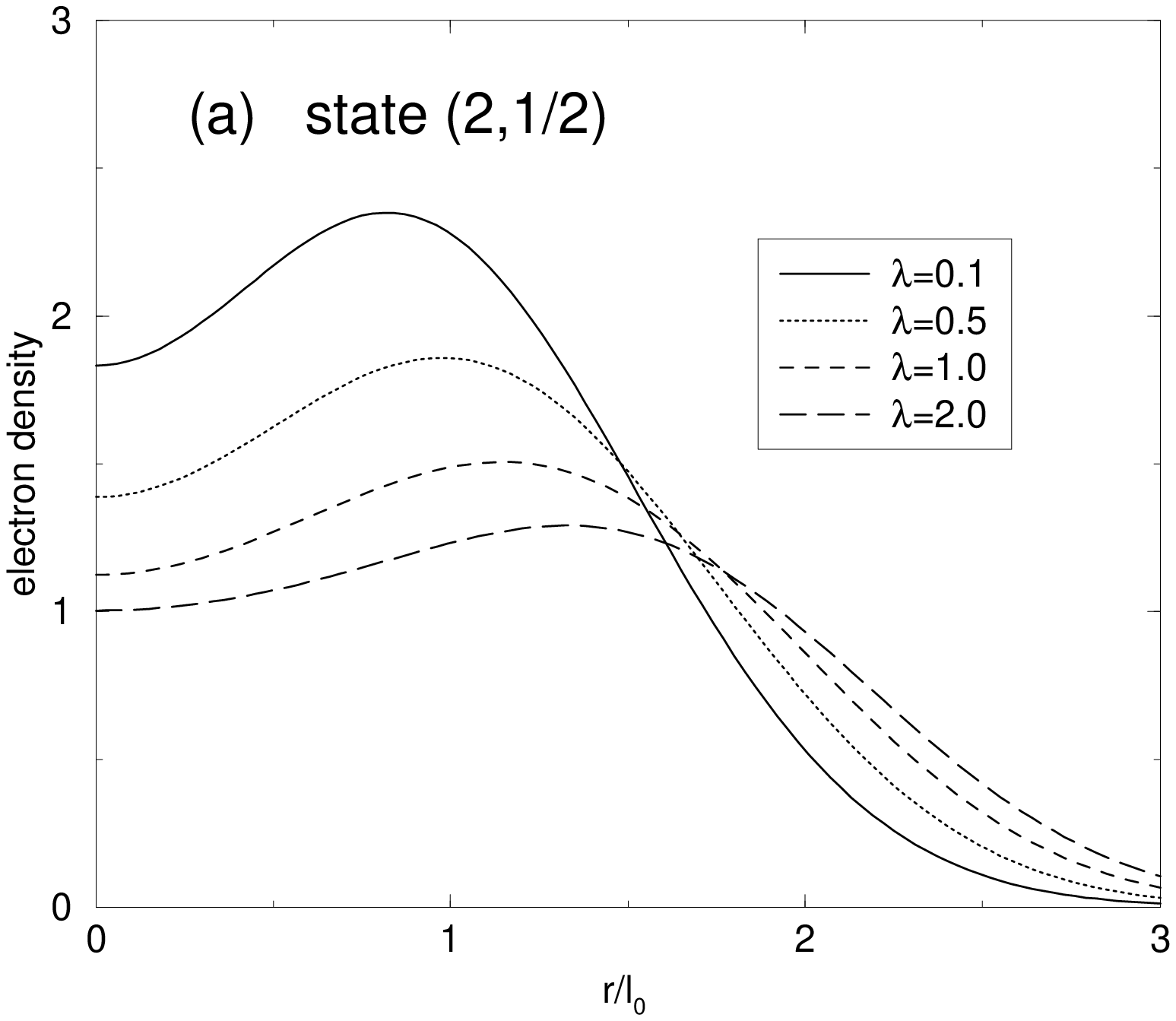,width=7.5cm}\hskip5mm\epsfig{file=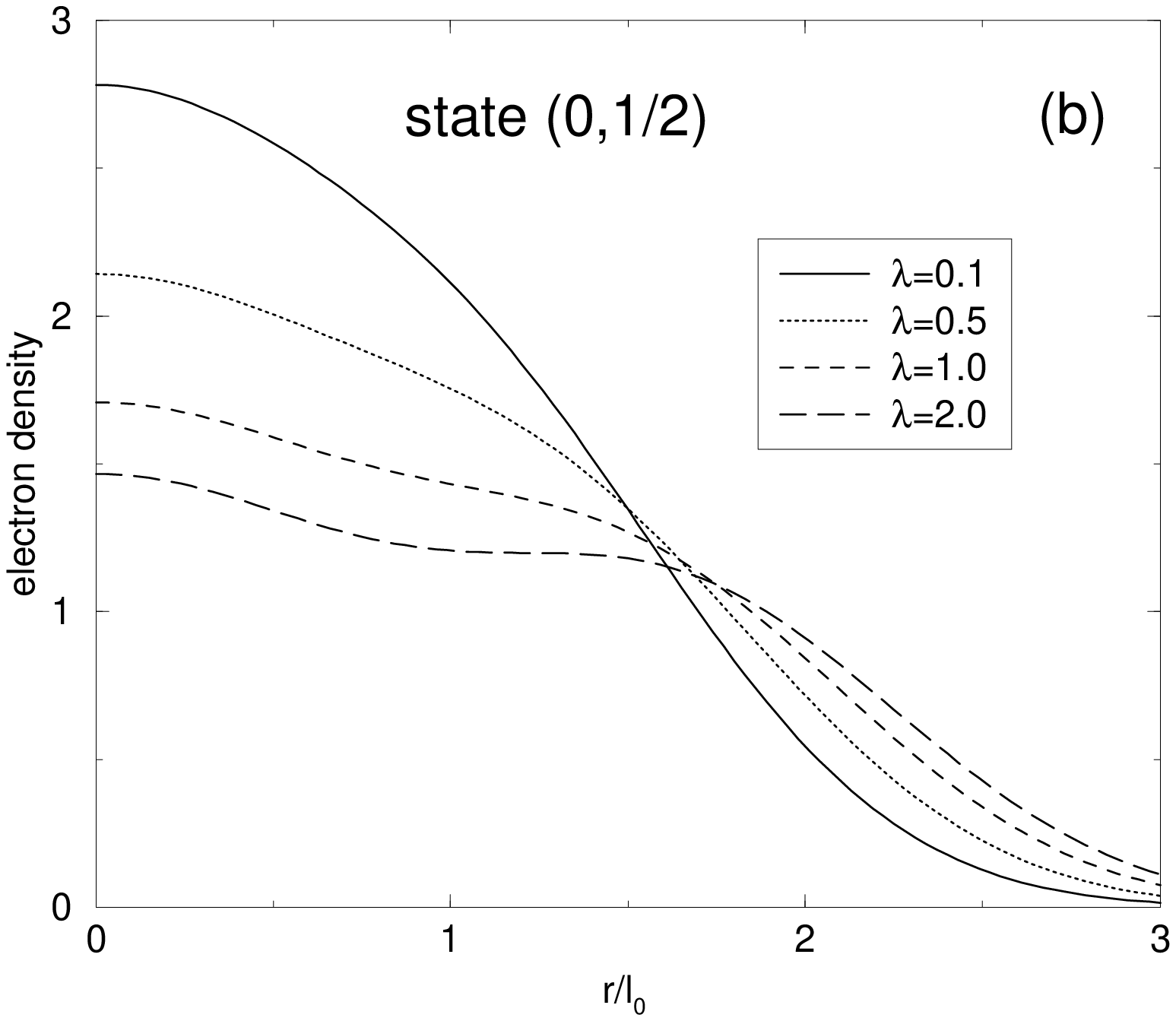,width=7.5cm}
\caption{The density of electrons in the ground (a) and the first excited (b) states at a few values of the interaction parameter $\lambda$.}
\label{7el-dens}
\end{figure}

{\sl Quantum dots.} Figure \ref{7el-energy} shows the two lowest energy levels of a 7-electron quantum dot as a function of the interaction parameter $\lambda$. In Figure \ref{7el-dens} the density of electrons in these states is analyzed at a few values of $\lambda$. One sees that in the first excited state, Fig. \ref{7el-dens}b, electrons behave as non- or weakly-interacting particles: they are mostly concentrated near the origin in spite of the Coulomb repulsion, and their total density is a smooth, almost monotonous function of $r$. This state is of a ``liquid'' (non-correlated) type. On the contrary, in the ground state, Fig. \ref{7el-dens}a, electrons are correlated: they avoid the origin due to the Coulomb repulsion, and the total electron density has a minimum at $r=0$. This state is of a ``solid'' (correlated) type. The state resembling the classical shell configuration [12] 
with one electron in the center and six electrons around, in the corners of a hexagon (maxima of the quantum-mechanical density at $r=0$ and at the radius of the shell) can be also observed in calculations, but in the completely spin-polarized state ($S_{tot}=7/2$), which is not the ground state. 

The ``solid''-type (correlated) state $(2,1/2)$ has the lower energy in the whole (studied) range of $\lambda$. The gap between the ground and the first excited states is however very small and does not exceed 0.33\%, Fig. \ref{7el-energy}b. This corresponds to a few-K temperature scale at typical parameters of GaAs dots ($\hbar\omega_0\approx 3$ meV). Thus, being at $T=0$ in the ``solid''-type ground state with essentially non-uniform density, the system ``melts'' to a ``liquid''-like state with a smoother quasi-uniform density at temperatures of order of a few K. No such ``melting'' is observed in the ground state (at $T=0$) at varying interaction parameter $\lambda$.

\begin{figure}[t]
\centering
\epsfig{file=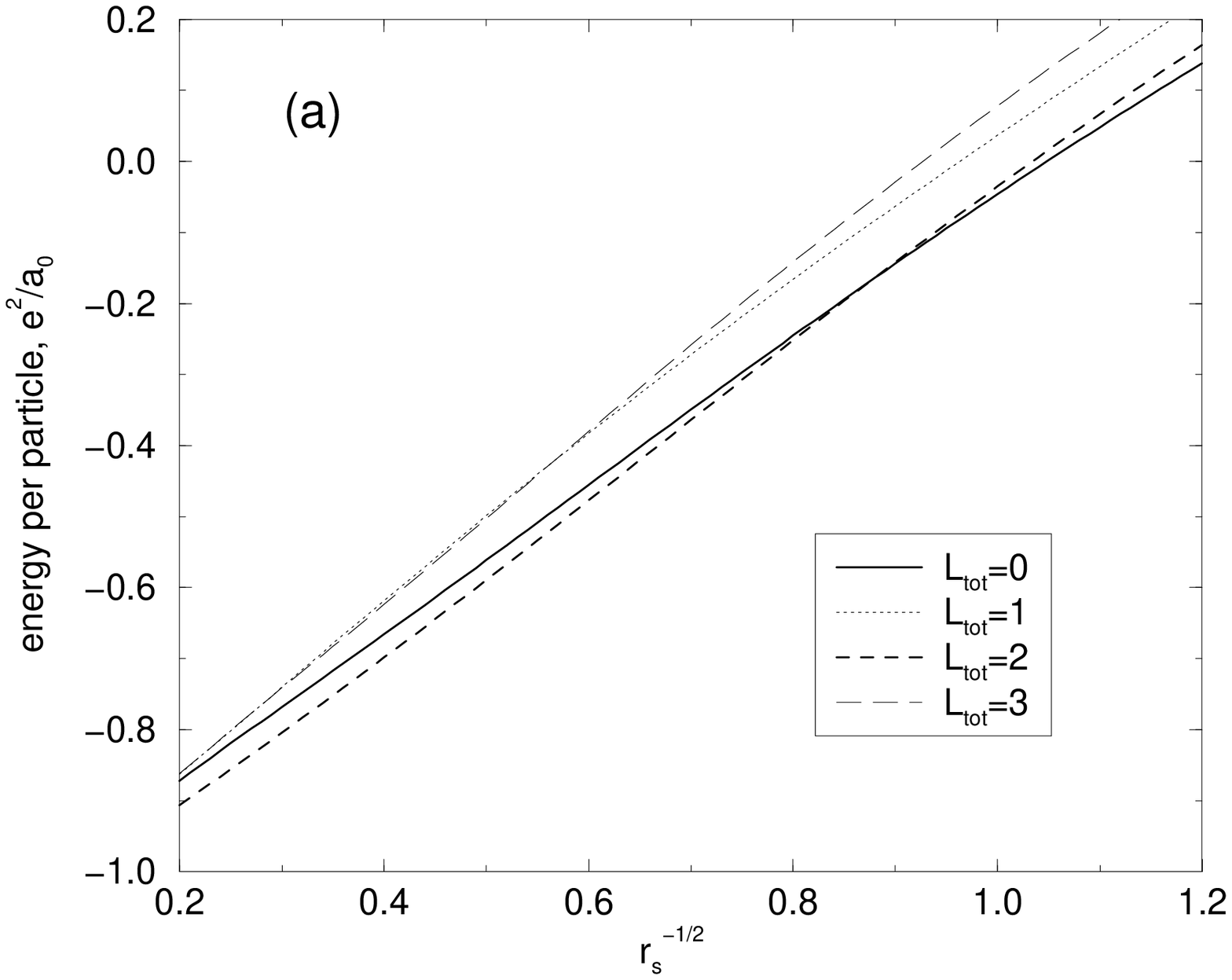,width=7.5cm}\hskip5mm\epsfig{file=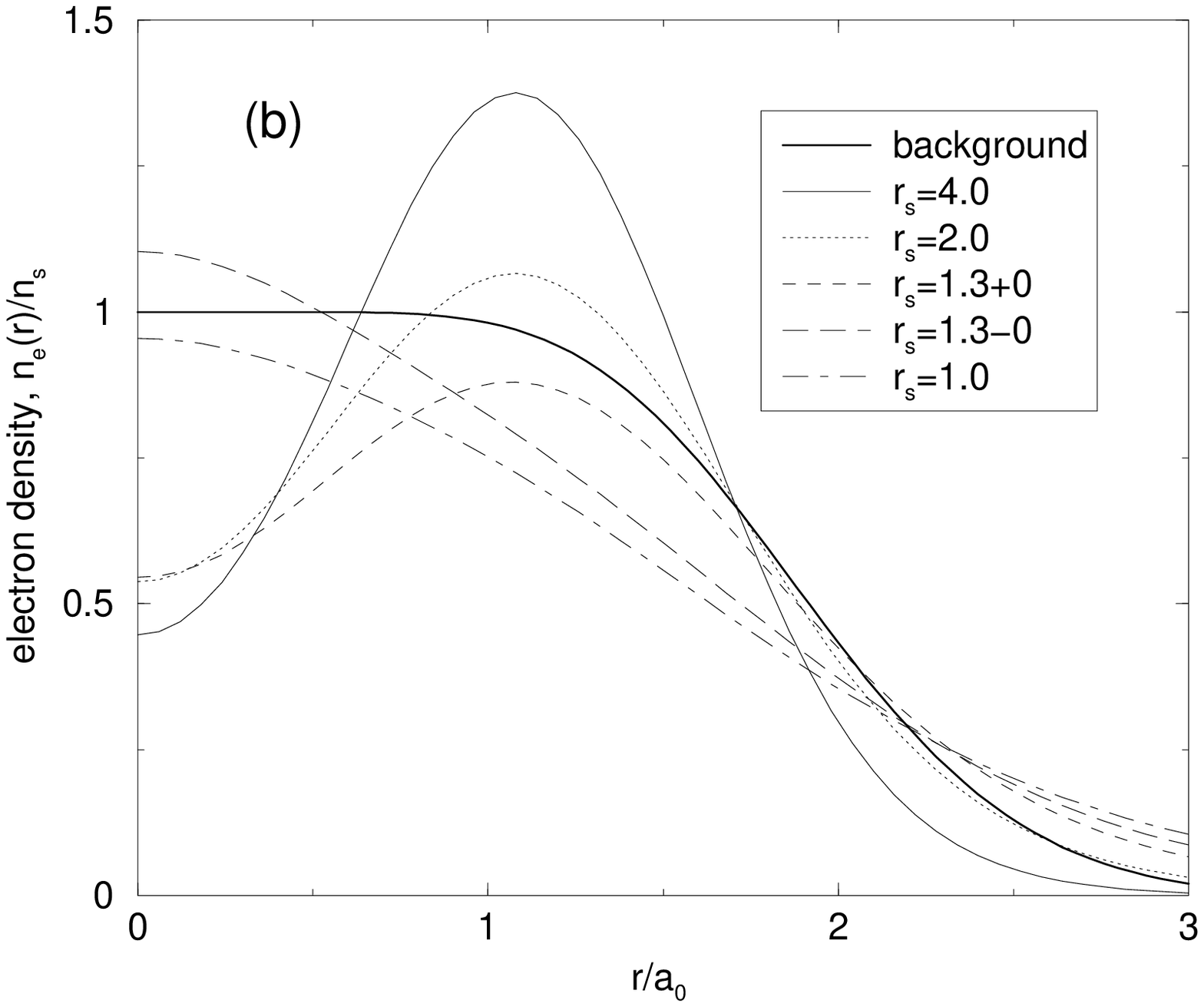,width=7.5cm}
\caption{Four-electron spin-polarized quantum disk: (a) energy per particle, in units $e^2/a_0$, vs $r_s^{-1/2}$, and (b) background and electron densities vs $r/a_0$, at a number of $r_s$-points near the transition at $r_s=1.3$. }
\label{N4}
\end{figure}

{\sl Quantum disks.} In quantum disks the situation is {\it qualitatively} different. Figure \ref{N4} shows results for a spin-polarized ($S_{tot}=2$) 4-electron quantum disk. At $r_s>r_c\approx 1.3$ the ground state has the total angular momentum $L_{tot}=2$ and the ``solid''-like (correlated) form of the density, while at smaller $r_s<r_c$ the ground-state angular momentum is $L_{tot}=0$ and the density has a ``liquid''-type (uncorrelated) form. The behavior of the density changes very sharply at the transition point, and contrary to dots, the transition occurs in the ground state (this effect {\it is not the case} in a similar, 4-electron spin-polarized, parabolic quantum dot, where the ground-state angular momentum is $L_{tot}=2$ at all values of the interaction parameter). Qualitatively, the transition appears very similarly to that in the infinite 2DES [9], 
but the critical point $r_c$ in the small disk lies at substantially higher density. The question, how the critical parameter $r_c$ depends on the number of particles $N$ in the spin-polarized and non-spin-polarized ground states, calls for further investigations.

\mysection{Summary} I have presented results of an exact-diagonalization study of the ground and excited states of few-electron quantum dots and disks. The quantum-disk model is characterized by an essentially non-parabolic confining potential, and directly corresponds, in the thermodynamic limit, to a realistic macroscopic sample with the uniform background density. I have shown that in the more adequate quantum-disk model the liquid-solid transition in the ground state is similar to the transition in the infinite 2DES and qualitatively differs from that in quantum dots.

The work was supported by the Sonderforschungsbereich 
484, University of Augsburg.

\mysection{References} 
\newlength{\mydescr}
\settowidth{\mydescr}{[1]}
\begin{list}{}%
{\setlength{\labelwidth}{\mydescr}%
\setlength{\leftmargin}{\parindent}%
\setlength{\topsep}{0pt}%
\setlength{\itemsep}{0pt}}
\item[\textrm{[1]}]
U.~Merkt, J.~Huser, and M.~Wagner, Phys. Rev. B {\bf 43}, 7320 (1991).

\item[\textrm{[2]}]
T.~Ezaki, N.~Mori, and C.~Hamaguchi, Phys. Rev. B {\bf 56}, 6428 (1997).

\item[\textrm{[3]}]
S.~M.~Reimann, M.~Koskinen, and M.~Manninen, Phys. Rev. B {\bf 62}, 8108 (2000). 

\item[\textrm{[4]}]
R.~Egger, W. H\"ausler, C.~H.~Mak, and H.~Grabert, Phys. Rev. Lett. {\bf 82}, 3320 (1999); {\bf 83}, 462(E) (1999)

\item[\textrm{[5]}]
F.~Pederiva, C.~J.~Umrigar, and E.~Lipparini, Phys. Rev. B {\bf 62}, 8120 (2000). 

\item[\textrm{[6]}]
B.~Reusch, W. H\"ausler, and H.~Grabert, Phys. Rev. B {\bf 63}, 113313 (2001).

\item[\textrm{[7]}]
A.~V.~Filinov, M.~Bonitz, and Yu.~E.~Lozovik, Phys. Rev. Lett. {\bf 86}, 3851 (2001).

\item[\textrm{[8]}]
C.~Yannouleas and U. Landman, Phys. Rev. Lett. {\bf 82}, 5325 (1999).

\item[\textrm{[9]}]
B.~Tanatar and D.~M.~Ceperley, Phys. Rev. B {\bf 39}, 5005 (1989).

\item[\textrm{[10]}]
S.~T.~Chui and B.~Tanatar, Phys. Rev. Lett. {\bf 74}, 458 (1995).

\item[\textrm{[11]}]
E.~Abrahams, S.~V.~Kravchenko, and M.~P.~Sarachik, Rev. Mod. Phys. {\bf 73}, 251 (2001).

\item[\textrm{[12]}]
F.~Bolton and U.~R\"ossler, Superlatt. Microstruct. {\bf 13}, 139 (1993).

\end{list} 
\end{document}